\documentclass{article}
\usepackage{spconf,amsmath,graphicx}
\usepackage{amsfonts}
\usepackage{ amssymb }

\usepackage{enumitem}
\usepackage{soul}
\usepackage{xcolor}
\setlist{nosep, leftmargin=14pt}
\usepackage{graphicx} 
\usepackage{caption,subcaption}

\usepackage{mwe} 

\title{Predicting Hypoxia in Brain Tumors from Multiparametric MRI}

\name{ Daniele Perlo$^{1}$ \qquad Georgia Kanli$^{1,2}$ \qquad Selma Boudissa$^{1,2}$ \qquad Olivier Keunen$^{1,2}$ }

\address{$^{1}$ Translational Radiomics, Luxembourg Institute of Health, Luxembourg  \\ 
        $^{2}$ In-Vivo Imaging Platform, Luxembourg Institute of Health, Luxembourg}

\begin{document}
\maketitle

\begin{abstract}

This research paper presents a novel approach to the prediction of hypoxia in brain tumors, using multi-parametric Magnetic Resonance Imaging (MRI). Hypoxia, a condition characterized by low oxygen levels, is a common feature of malignant brain tumors associated with poor prognosis. Fluoromisonidazole Positron Emission Tomography (FMISO PET) is a well-established method for detecting hypoxia in vivo, but it is expensive and not widely available.

Our study proposes the use of MRI, a more accessible and cost-effective imaging modality, to predict FMISO PET signals. We investigate deep learning models (DL) trained on the ACRIN 6684 dataset, a resource that contains paired MRI and FMISO PET images from patients with brain tumors. Our trained models effectively learn the complex relationships between the MRI features and the corresponding FMISO PET signals, thereby enabling the prediction of hypoxia from MRI scans alone.

The results show a strong correlation between the predicted and actual FMISO PET signals, with an overall PSNR score above 29.6 and a SSIM score greater than 0.94, confirming MRI as a promising option for hypoxia prediction in brain tumors. This approach could significantly improve the accessibility of hypoxia detection in clinical settings, with the potential for more timely and targeted treatments.

\end{abstract}


\section{Introduction}

Fluoromisonidazole (FMISO) is a Positron Emission Tomography (PET) tracer that is commonly used for imaging hypoxic conditions in cells. Hypoxia, or low oxygen level, is prevalent in many high grade solid tumors including head-and-neck cancers, glioblastoma, gastrointestinal tumors, lung and breast cancers. Hypoxic tumors are generally associated with an unfavorable prognosis due to a higher resistance to radiotherapy and increased risk of recurrence after surgery or chemotherapy~\cite{Kniess2023}.

FMISO consists of 2-nitroimidazole molecule labeled with the radioisotope Fluorine-18. It is selectively taken up by hypoxic cells and retained, while in normoxic (normal oxygen level) tissues, FMISO is reduced and reoxidized, allowing it to diffuse back out of the cell~\cite{Kniess2023}. This property makes FMISO a valuable tool for identifying and quantifying hypoxia in tissues, providing a non-invasive method to assess oxygen demand in cancer cells and its potential impact on radiotherapy and chemotherapy~\cite{Kniess2023}. This information is key for the therapeutic management of patients, as it has been found that for hypoxic tumors, a 2.5- to 3-fold higher radiotherapy dose is necessary to achieve the same cytotoxic effect~\cite{Watanabe2019}.
FMISO has thus established itself as an important tracer in the field of oncology, providing critical insight into tumor hypoxia that can guide treatment strategies and improve patient outcomes. 

Magnetic Resonance Imaging (MRI) is another modality used in the context of brain tumors, in which different sequences provides complementary contrasts useful to visualize various aspects of healthy and pathological tissues. The sequences most commonly used include T1-weighted (T1w), T1-weighted after the injection of a Gadolinium-based contrast agent (T1Gd), T2-weighted (T2w), and Fluid-Attenuated Inversion Recovery (FLAIR). Commonly referred to as 'anatomical' MRI, these sequences together provide insight into the different compartments of the tumor. T1w and T1Gd sequences are notably used for tumor detection by highlighting the regions of the brain where the blood-brain-barrier is disrupted, informing on the proliferating compartment and necrotic sections of the tumors. T2w images provide complementary information on the edema surrounding the tumor, inflammation, blood products and calcification within the tumor. FLAIR provides similar information to a T2w image, but with the signal of the cerebrospinal fluid suppressed, making it possible to better visualize the tumor boundary in the vicinity of ventricles.

Other less frequently used sequences have also been proposed to image tissue oxygenation in the tumors. They include T2* weighted (T2*) and Blood Oxygen Level Dependent (BOLD) sequences, as well as perfusion techniques such as Dynamic Susceptibility Contrast (DSC) and Dynamic Contrast Enhanced (DCE) MRI~\cite{DAlonzo2021,Gouel2023,Lee2014}. While each of these techniques has its merits and does provide signal that is influenced by the oxygenation status of the tissue, none provides a true image of tissue hypoxia. In the present study, we propose that the combination of classical anatomical sequences contains signatures of healthy and pathological tissues that can be used to predict hypoxia.

\section{Related Work}

Image-to-Image Translation is a task in computer vision and machine learning where the goal is to learn a mapping between an input image and an output image. Multiple solutions like CycleGAN~\cite{Zhu2017-ic} and Pix2Pix~\cite{Isola2016-ze} have been proposed to handle this task.
MRI-to-MRI translation, also known as Image Modality Translation (IMT), involves translating images from one MR\st{I} sequence to another, for instance using DL models like Conditional Generative Adversarial Networks (CGANs)~\cite{Yang2020-hf}. IMT is mostly used as support for other imaging  tasks, like registration or segmentation~\cite{Yang2020-hf} or for data augmentation.
With anatomical MRI translation, IMT uses structural information of the brain in order to predict a different contrast, such as from a T1w image to a FLAIR image, but this translation typically does not involve extracting features related to tissue state. With anatomical MRI-to-PET image translation, the task grows in complexity as the models have to learn a mapping between structural information and tissue function.

The majority of the literature on the prediction of PET images from MRI relates to the generation of Fluorodeoxyglucose (FDG) PET images, since this imaging modality is the most commonly used in the clinic. FDG is an analog of glucose, and cancer cells tend to have an increased absorption of it because of their high energy demanding proliferating state.
A first cross-modality attempt to predict FDG PET images from CT in liver lesions used a combination of fully connected and conditioned GAN networks~\cite{Ben2018}, and more recently U-Net based models~\cite{Bhat2022}. From 2020, GAN approaches have been proposed, that can learn to generate synthetic 2 dimensional low-resolution FDG images directly to populate synthetic datasets~\cite{Islam2020,Abazari2022}.
The most common data source for PET generation is a large database of patients with Alzheimer's disease called ADNI~\cite{Petersen2009-ww}. Other datasets with cross-modality imaging (MRI, CT and PET) have also been made publicly available, including OASIS~\cite{LaMontagne2019} and CERMEP-IDB-MRXFDG~\cite{Merida2021}.
A 2D MRI to MRI cross-modality prediction method is then proposed with ResViT, a pix2pix method with adversarial training (p2p GAN) enhanced by the use of transformers blocks, showing extremely accurate MRI predictions~\cite{Dalmaz2022}.
In 2022, MRI-FDG PET synthesis task received further attention, with T1 images chosen as input to models such as U-PET~\cite{Kollovieh2022}, BPGAN~\cite{Zhang2022} and output in E-GAN~\cite{Bazangani2022}. 
In the last year, multiple T1 to FDG prediction methods are again proposed~\cite{Zadeh2023}, relying on U-Net variants~\cite{Rajagopal2023}, diffusion models~\cite{Xie2023} or GAN~\cite{Ouyang2023,Tu2024,Karimipourfard2023-gw}. As far as we know, DL-based PET prediction attempts in oncology for radiotracers other than FDG have been limited to the prediction of $^{11}C$-methionine PET~\cite{Takita2023} and the hypoxia-specialized tracer HX4~\cite{Traverso2022}.

While most of the proposed studies use FDG PET as target for PET signal prediction, the metabolic activity evidenced by FDG is of limited value to predict intra-tumor hypoxia. While there may be a correlation between FDG and FMISO uptake in some tumors, discordant results have been found in others, suggesting that hypoxia and high metabolic activity do not always coincide~\cite{Nehmeh2021}.
In the present paper we compared different DL approaches to generate PET images, and evaluated their ability to predict the FMISO PET signal from multi-parametric MRI. Given the limited size of our input dataset, we also evaluated if pretraining models to predict FDG signal can be beneficial for the prediction of FMISO. We evaluated a 3D adaptation of ResViT~\cite{Dalmaz2022} and compared it to other models commonly used in the litterature. We used multiple MRI channels, in order to exploit the unique MRI signature provided by the combination of complementary anatomical series and compared the performance achieved to predictions obtained from single channel T1-weighed imaging.

\section{Method}

\subsection{Dataset description}

\subsubsection{FMISO Dataset}

In order to access coupled pairs of MRI and FMISO PET images, we used the ACRIN~6684 dataset~\cite{ACRIN}. Created to study survival in patients diagnosed with glioblastoma multiforme (GBM),  ACRIN~6684 provides clinical data as well as matching MRI, PET and CT images. 
The cohort contains 51 cases from 47 patients, of which we limited ourselves to the cases where MR images were collected within 30 days form the PET acquisitions, to account for the typically high growth rate of GBM.

\subsubsection{FDG Dataset}

The versions of our models pretrained with FDG PET images made use of the open access OASIS-3 dataset~\cite{LaMontagne2019}. OASIS contains matching MRI and PET sequences from 1379 different subjects. This resource primarily assembled to study the early stages of Alzheimer’s Disease (AD) happens to contain a large fraction of healthy controls. We used T1-weighted, T2* and FLAIR MRI sequences for the prediction of FDG. The selection of matching MRI-FDG images is done with similar criteria to those used for the MRI-FMISO samples, with the searching window enlarged to 90 days to account for the high fraction of healthy brains and slow disease progression in early stage AD patients.

Despite the overall large size of the dataset, only 92 cases had both FDG and associated T1-weighted acquisitions. T2* and FLAIR images were also available in 42 and 11 cases respectively.

\subsection{Data Preprocessing}

\begin{figure}[htb!]
\centering
\includegraphics[width=\linewidth]{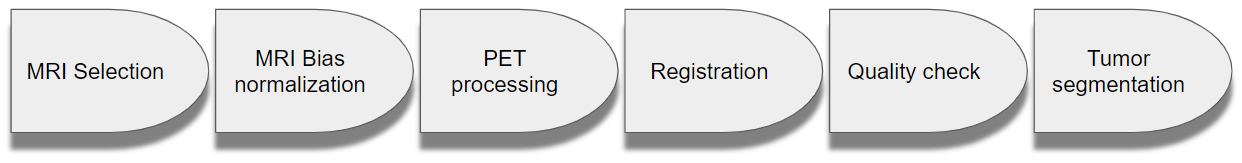}
\caption{Data preparation workflow for FMISO and FDG\\ images.}
\label{fig:pipeline}
\end{figure}

In both datasets, MRI and PET brain images are not registered because acquired on different machines and at different dates. Moreover a resampling of the brain volumes is needed to account for differences in number of scan sections, spatial resolution, scan orientation and head position.
DL solutions need a good data quality in order to be capable of extracting meaningful features from input images, therefore the registration process is crucial. In each area of the co-registered brains, each MRI protocols contributes complementary information on the underlying tissue. But co-registration of MRI and PET volumes also comes with specific challenges. While tools do exist for MRI sequences co-registration, segmentation and skull stripping, they are less suitable for PET images, where the anatomical structures and landmarks of the brain are often lacking.

We achieved satisfying brain volumes registration by implementing the following pipeline, illustrated in Fig 1:
\begin{enumerate}
\item{Selection of a reference MRI protocol:} The brain volume of the reference protocol is the one that is register to a pre-defined atlas. In our case we use a T2w atlas~\cite{Rohlfing2010-zm}. The reference is resampled to the atlas space and resolution. All other protocols and PET images are then registered to the reference. We find the T2-weighted sequence to be the optimal protocol to use for ACRIN~6684, because T2W images had been acquired at higher resolution than other protocols. 
\item{MRI Bias normalization:} Bias normalization is used to correct for intensity variations that can occur due to different scanners or parameters used during the MR image acquisition process~\cite{Sun2015}. N4 Bias normalization is used in our pipeline.
\item{PET processing:} PET static acquisition involves the collection of data over a fixed period of time after the injection of the radiotracer~\cite{K2013}. From the dynamic PET acquisitions available in the ACRIN 6684 and OASIS databases, we averaged the signal corresponding to the xx to xx time window after tracer injection for FMISO and xx to xx for FDG respectively.
The collected PET images are often noisy and show attenuation/reconstruction artifacts in the area surrounding the patient head. Therefore we calculate a binary mask to apply to the PET, in order to facilitate the registration process. The mask $M$ of a PET volume $p$ is calculated as:
\begin{equation}
  M(p) = Otsu(\phi(\psi(p)))
\end{equation}
where $\psi$ is a Gaussian smooth operator with $\sigma=1$, $\phi$ is a contrast enhancing operator with $\gamma=1$ and the $Otsu$ method finds the threshold value for the signal and then the binary mask. 
\item{Registration:} As main registration tool, we use the SPM MATLAB-based API~\cite{Friston2011-jz}, wrapped in our Python pipeline using Nypipe~\cite{gorgolewski_2016_50186}. SPM typically provides quality registration, but can fail if the input signal does not have a sufficient partial overlap with the reference. Therefore, a first coarse optional registration is applied by using the Greedy Registration tool~\cite{Menze2015-fb}. Optionally, the registration offered by Slicer3D~\cite{Fedorov2012-qe} can also be used in exceptionally difficult cases.
\item{Quality control:} Mutual Information and Dice Score are then computed to evaluate the quality of the registration. If the metrics for a pair registration fall below a given threshold, the case is discarded alltogether and not further retained for models training. 
\item{Tumor segmentation:} We use DeepMedic~\cite{Kamnitsas2017-rm} on T1Gd, T1, T2 and FLAIR images to segment the area of the brain tumor. Hypoxia signals are visible in FMISO PET in the peripheral region of the tumor, where the oxygen supply to the tissues is compromised.  
\end{enumerate} 

\subsection{Model}

We trained the ResViT model to capture spatial and multi-parametric features from the different MRI sequences. Each image volume is resized at the high resolution of $256\times256\times128$ pixels. The model is originally proposed to work with 2 dimensional multi-channel input, so every layer is substituted by its 3D counterpart, by mirroring each layer parameters on the new dimension.
Therefore, the original generator function $G(x)$ change its domain from $\mathbb{R}^3 \rightarrow \mathbb{R}^2$ to $\mathbb{R}^4 \rightarrow \mathbb{R}^3$, while the discriminator function $D(x)$ is redefined from $\mathbb{R}^2 \rightarrow \mathbb{R}$ to $\mathbb{R}^3 \rightarrow \mathbb{R}$ and LSGAN loss is applied. The embedding parameter in the Aggregate Residual Transformer modules of ResViT are also expanded to fit the encoder output. 

The ResViT loss function $\mathcal{L}$ is then enhanced by adding a penalty term Tumor Focus $\mathcal{L}_{focus}$ to increase the fidelity of the prediction around the tumor region:
\begin{equation}
  \mathcal{L} = \lambda_1 \mathcal{L}_{gan} + \lambda_2 \mathcal{L}_{L1} + \lambda_3 \mathcal{L}_{focus}
\end{equation}
where 
\begin{equation}
  \mathcal{L}_{focus}(G(x),y) = |M(G(x)) - M(p)|
\end{equation}
A dilation operator is applied on the tumor region, with a 3 dimensional kernel, for 3 iterations, in order to include the surrounding tissues around the main lesion.

We also applied simple data augmentation as horizontal flip (right t and left side of the brain) on both input MRI sequences and PET images with probability $p=0.5$. A random bias field augmentation, with polynomials degree 3, is applied independently on each input channel with $p=0.5$, to increase the model robustness.

\begin{figure*}[ht!]
\centering
\includegraphics[width=\linewidth]{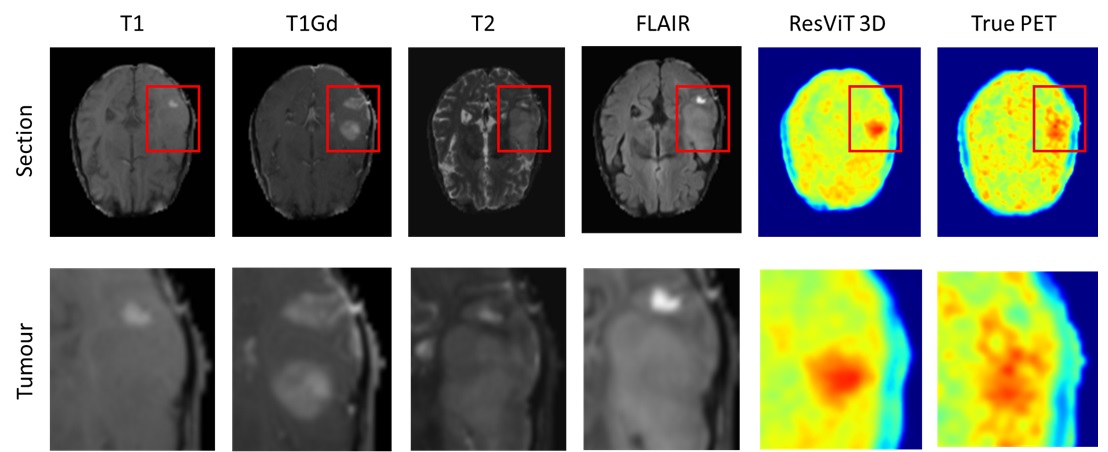}
\caption{Axial view of cancer region FMISO signal reconstruction from our 3D ResViT trained with Tumor Focus.}
\label{fig:fmiso_zoom}
\end{figure*}

\begin{table*}[th!]
\centering
\caption{Comparing the results between strategies and architectures. $\lambda_3$ for the Tumor Focus is set to 100. }
\label{tab:resultsT}
\begin{tabular}{|c||c|c|c|c|}
\hline
& \multicolumn{2}{c|}{Brain Volume}& \multicolumn{2}{c|}{Tumor Bounding Box} \\
\hline
\textbf{Model - 4 MRI Channels}                               &PSNR &SSIM&PSNR                    & SSIM                   \\
\hline
Pix2Pix (2D)~\cite{Isola2016-ze}                                & 29.499$\pm$1.924                & 0.900$\pm$0.062                & 21.460$\pm$2.776                & 0.728$\pm$0.092                \\
ResViT (2D)~\cite{Dalmaz2022}                                 & \textbf{30.190$\pm$2.123}                & \textbf{0.926$\pm$0.047 }               & 21.460$\pm$3.266                & 0.782$\pm$0.090                \\
Residual U-Net (3D)~\cite{Rajagopal2023}                       & 26.835$\pm$0.696                & 0.893$\pm$0.033                & 19.590$\pm$3.000                & 0.609$\pm$0.126                \\
ResViT (3D)                                 & 29.578$\pm$1.032                & 0.931$\pm$0.021                & \textbf{23.653$\pm$2.824}                & \textbf{0.816$\pm$0.042}                \\
\hline
\textbf{MRI-FDG Pretrained}                                & \multicolumn{4}{c|}{}\\
\hline
Residual U-Net (3D)~\cite{Rajagopal2023}                         & 29.198$\pm$1.015                & 0.937$\pm$0.021                & 21.628$\pm$3.027                & 0.748$\pm$0.063                \\
ResViT (3D)                                 & \textbf{29.756$\pm$1.038 }               & \textbf{0.944$\pm$0.013}                & 22.946$\pm$2.092                & 0.795$\pm$0.035                \\
ResViT (3D)  with Tumor Focus               & 29.615$\pm$1.099               & 0.938$\pm$0.014                & \textbf{24.042$\pm$2.632}                & \textbf{0.795$\pm$0.032}                \\
\hline
\textbf{Single Channels}                             & \multicolumn{4}{c|}{} \\
\hline
ResViT (3D) - T2    & 28.900$\pm$1.271                & 0.883$\pm$0.051                & 23.417$\pm$2.912                & 0.786$\pm$0.056                \\
ResViT (3D) - T1    & 28.879$\pm$0.757                & 0.938$\pm$0.020                & \textbf{23.469$\pm$2.580}                & \textbf{0.787$\pm$0.034}                \\
ResViT (3D) - T1Gd  & 28.917$\pm$1.183                & 0.935$\pm$0.026                & 23.020$\pm$3.558                & 0.763$\pm$0.067                \\
ResViT (3D) - FLAIR & \textbf{29.300$\pm$1.303}                & \textbf{0.942$\pm$0.007 }               & 21.754$\pm$2.426                & 0.757$\pm$0.038               \\
\hline
\end{tabular}
\end{table*}

\section{Results}

Results are obtained by training each solution for 50 epochs with fixed learning rate, followed by 20 epochs with linear learning rate decay. An initial learning rate of 0.0001 provided the best results for both ResViT and compared architectures. All training jobs are done on a server equipped with NVIDIA RTX A6000 GPU cards. All inputs are masked with a brain mask provided by applying skull stripping on the T2 images, in order to not consider bones and areas outside the brain tissues. The test set represents 10\% of the ACRIN 6684 dataset, making sure that exams from a given patient do not fall in the same dataset split to avoid training bias.

Fig.~\ref{fig:fmiso_zoom} illustrates the predictions of FMISO signal achieved with the 3D ResViT model and multi-channels MRI input, in comparison to the true FMISO PET acquisitions. Table~\ref{tab:resultsT} shows that the performance achieved by the 2D models in reconstructing the FMISO PET signal in the tumor regions was lower than with 3D models. 2D models achieved better results in predicting the overall slice image, with a PSNR of 30.190 and a SSIM of 0.926, but in the tumor area where the FMISO signal is most clinically informative, the 3D ResViT achieved the best performance wit a PSNR of 23.653 and a SSIM of 0.816. Prediction of FMISO PET signal using Cycle GAN did not produced meaningful results in our study, as the method is better suited for 1-to-1 prediction models and less adapted to our many-to-1 application. Our 3D version of ResViT outperformed 3D Residual U-Net~\cite{Rajagopal2023} and Pix2Pix models, both considering the training from scratch and from a model pretrained for FDG prediction. The Tumor Focus penalty did not impact the performances in a significant way when assessing the complete reconstructed FMISO brain, but helps to reach a better PSNR score on the bounding box surrounding the tumour area. Using a network pretrained for FDG prediction from MRI only provided a marginal gain.

We also examined the relative contribution of each MRI channel to the quality of the FMISO signal prediction. Feeding the models with a single MRI channel as inputs already provided good overall performance, with the only exception of FLAIR. Providing FLAIR as input, the model perform well in predicting the overall FMISO signal in the brain, but performed poorly in the brain region, where the precision of the prediction matters most. The multi-channels model achieved the best combined performance in overall image quality and tumor area.

\section{Conclusions}

Predicting hypoxia in brain tumors from anatomical MRI series is a challenging task. Single channel models provide a first approximation but lack specificity in the clinically most relevant areas such as tumors. Multi-channel models improve prediction, notably in the eloquent regions, despite the added challenges caused by images co-registration and standardization. In future, we will inestigate if models that combine anatomical series with functional and physiological series such as diffusion, perfusion, T2* or BOLD images can further improve predictions by extracting features that can relate more closely to hypoxia.

Further research is also needed to validate these findings in larger patient cohorts and across different types of brain tumors. As for most AI models developed for medical applications, datasets can be very limited in the number of significant samples available. Moreover, often all the protocols available in public access datasets are proposed at different resolution, often in term of number of image slices. By applying voxel interpolation to resizing the images to a shared fixed resolution, the preprocessing step can introduce small artifacts. In order to build a more accurate predictor, we are now collecting pre-clinical MRI-FMISO scan pairs from our laboratories. Other image-to-image methods, such as the recently introduced image conditioned diffusion models may also be able to improve FMISO signal generation.

Nonetheless, this study represents a first significant step in the non-invasive assessment of tumor hypoxia from multi-parametric MRI. This could ultimately contribute to improved patient outcomes in brain tumor management, providing support to healthcare professionals in achieving more precise diagnoses and formulating more effective treatment plans.

\section{Acknowledgment}
This work was supported by the Fonds National de la Recherche Scientifique-FNRS under the  Télévie 7.4554.22 grant 

\bibliographystyle{ieeetr}
\bibliography{bib}

\begin{thebibliography}{10}

\bibitem{Kniess2023}
T.~Kniess, J.~Zessin, P.~M{\"a}ding, M.~Kuchar, O.~Kiss, and K.~Kopka, ``Synthesis of {[18F]FMISO}, a hypoxia-specific imaging probe for {PET}, an overview from a radiochemist's perspective,'' {\em EJNMMI Radiopharmacy and Chemistry}, vol.~8, p.~5, Mar 2023.

\bibitem{Watanabe2019}
S.~Watanabe, T.~Inoue, S.~Okamoto, K.~Magota, A.~Takayanagi, J.~Sakakibara-Konishi, N.~Katoh, K.~Hirata, O.~Manabe, T.~Toyonaga, Y.~Kuge, H.~Shirato, N.~Tamaki, and T.~Shiga, ``Combination of {FDG-PET} and {FMISO-PET} as a treatment strategy for patients undergoing early-stage {NSCLC} stereotactic radiotherapy,'' {\em EJNMMI Research}, vol.~9, p.~104, Dec 2019.

\bibitem{DAlonzo2021}
R.~A. D’Alonzo, S.~Gill, P.~Rowshanfarzad, S.~Keam, K.~M. MacKinnon, A.~M. Cook, and M.~A. Ebert, ``In vivo noninvasive preclinical tumor hypoxia imaging methods: a review,'' {\em International Journal of Radiation Biology}, vol.~97, p.~593–631, Apr. 2021.

\bibitem{Gouel2023}
P.~Gouel, P.~Decazes, P.~Vera, I.~Gardin, S.~Thureau, and P.~Bohn, ``Advances in pet and mri imaging of tumor hypoxia,'' {\em Frontiers in Medicine}, vol.~10, Feb. 2023.

\bibitem{Lee2014}
C.-T. Lee, M.-K. Boss, and M.~W. Dewhirst, ``Imaging tumor hypoxia to advance radiation oncology,'' {\em Antioxidants \& Redox Signaling}, vol.~21, p.~313–337, July 2014.

\bibitem{Zhu2017-ic}
J.-Y. Zhu, T.~Park, P.~Isola, and A.~A. Efros, ``Unpaired image-to-image translation using cycle-consistent adversarial networks,'' Mar. 2017.

\bibitem{Isola2016-ze}
P.~Isola, J.-Y. Zhu, T.~Zhou, and A.~A. Efros, ``Image-to-image translation with conditional adversarial networks,'' Nov. 2016.

\bibitem{Yang2020-hf}
Q.~Yang, N.~Li, Z.~Zhao, X.~Fan, E.~I.-C. Chang, and Y.~Xu, ``{MRI} cross-modality image-to-image translation,'' {\em Sci. Rep.}, vol.~10, p.~3753, Feb. 2020.

\bibitem{Ben2018}
A.~Ben-Cohen, E.~Klang, S.~P. Raskin, S.~Soffer, S.~Ben-Haim, E.~Konen, M.~M. Amitai, and H.~Greenspan, ``Cross-modality synthesis from {CT} to {PET} using {FCN} and {GAN} networks for improved automated lesion detection,'' 2 2018.

\bibitem{Bhat2022}
S.~Bhat, T.~Arsenault, A.~Baydoun, L.~Bailey, A.~Amini, B.~George, K.~Nam, G.~Saieed, R.~A. Zeidane, J.~U. Heo, R.~Muzic, T.~Biswas, and T.~Podder, ``Synthetic {FDG}-positron emission tomography images for patients with non-small cell lung cancer: A deep learning-based approach using computed tomography images,'' {\em International Journal of Radiation Oncology*Biology*Physics}, vol.~114, pp.~e127--e128, 11 2022.

\bibitem{Islam2020}
J.~Islam and Y.~Zhang, ``{GAN}-based synthetic brain {PET} image generation,'' {\em Brain Informatics}, vol.~7, p.~3, 12 2020.

\bibitem{Abazari2022}
M.~A. Abazari, M.~Soltani, F.~M. Kashkooli, and K.~Raahemifar, ``Synthetic {18F-FDG PET} image generation using a combination of biomathematical modeling and machine learning,'' {\em Cancers}, vol.~14, p.~2786, 6 2022.

\bibitem{Petersen2009-ww}
R.~C. Petersen, P.~S. Aisen, L.~A. Beckett, M.~C. Donohue, A.~C. Gamst, D.~J. Harvey, C.~R. Jack, Jr, W.~J. Jagust, L.~M. Shaw, A.~W. Toga, J.~Q. Trojanowski, and M.~W. Weiner, ``Alzheimer's disease neuroimaging initiative ({ADNI}): clinical characterization,'' {\em Neurology}, vol.~74, pp.~201--209, Dec. 2009.

\bibitem{LaMontagne2019}
P.~J. LaMontagne, T.~L. Benzinger, J.~C. Morris, S.~Keefe, R.~Hornbeck, C.~Xiong, E.~Grant, J.~Hassenstab, K.~Moulder, A.~G. Vlassenko, M.~E. Raichle, C.~Cruchaga, and D.~Marcus, ``{OASIS-3}: Longitudinal neuroimaging, clinical, and cognitive dataset for normal aging and alzheimer disease,'' Dec. 2019.

\bibitem{Merida2021}
I.~Mérida, J.~Jung, S.~Bouvard, D.~L. Bars, S.~Lancelot, F.~Lavenne, C.~Bouillot, J.~Redouté, A.~Hammers, and N.~Costes, ``{CERMEP-IDB-MRXFDG}: a database of 37 normal adult human brain {[18F]FDG PET, T1 and FLAIR MRI, and CT} images available for research,'' {\em EJNMMI Research}, vol.~11, p.~91, 12 2021.

\bibitem{Dalmaz2022}
O.~Dalmaz, M.~Yurt, and T.~Cukur, ``Resvit: Residual vision transformers for multimodal medical image synthesis,'' {\em IEEE Transactions on Medical Imaging}, vol.~41, pp.~2598--2614, 10 2022.

\bibitem{Kollovieh2022}
M.~Kollovieh, M.~Keicher, S.~Wunderlich, H.~Burwinkel, T.~Wendler, and N.~Navab, ``{U-PET}: Mri-based dementia detection with joint generation of synthetic {FDG-PET} images,'' 6 2022.

\bibitem{Zhang2022}
J.~Zhang, X.~He, L.~Qing, F.~Gao, and B.~Wang, ``{BPGAN}: Brain {PET} synthesis from mri using generative adversarial network for multi-modal alzheimer’s disease diagnosis,'' {\em Computer Methods and Programs in Biomedicine}, vol.~217, p.~106676, 4 2022.

\bibitem{Bazangani2022}
F.~Bazangani, F.~J.~P. Richard, B.~Ghattas, and E.~Guedj, ``{FDG-PET} to t1 weighted mri translation with 3d elicit generative adversarial network (e-gan),'' {\em Sensors}, vol.~22, p.~4640, 6 2022.

\bibitem{Zadeh2023}
F.~S. Zadeh, S.~Molani, M.~Orouskhani, M.~Rezaei, M.~Shafiei, and H.~Abbasi, ``Generative adversarial networks for brain images synthesis: A review,'' 5 2023.

\bibitem{Rajagopal2023}
A.~Rajagopal, Y.~Natsuaki, K.~Wangerin, M.~Hamdi, H.~An, J.~J. Sunderland, R.~Laforest, P.~E. Kinahan, P.~E.~Z. Larson, and T.~A. Hope, ``Synthetic {PET} via domain translation of {3-D MRI},'' {\em IEEE Transactions on Radiation and Plasma Medical Sciences}, vol.~7, pp.~333--343, 4 2023.

\bibitem{Xie2023}
T.~Xie, C.~Cao, Z.~Cui, Y.~Guo, C.~Wu, X.~Wang, Q.~Li, Z.~Hu, T.~Sun, Z.~Sang, Y.~Zhou, Y.~Zhu, D.~Liang, Q.~Jin, G.~Chen, and H.~Wang, ``Synthesizing {PET} images from high-field and ultra-high-field mr images using joint diffusion attention model,'' 5 2023.

\bibitem{Ouyang2023}
J.~Ouyang, K.~T. Chen, R.~D. Armindo, G.~A. Davidzon, K.~E. Hawk, F.~Moradi, J.~Rosenberg, E.~Lan, H.~Zhang, and G.~Zaharchuk, ``Predicting {FDG‐PET} images from multi‐contrast {MRI} using deep learning in patients with brain neoplasms,'' {\em Journal of Magnetic Resonance Imaging}, 6 2023.

\bibitem{Tu2024}
Y.~Tu, S.~Lin, J.~Qiao, Y.~Zhuang, Z.~Wang, and D.~Wang, ``Multimodal fusion diagnosis of alzheimer’s disease based on {FDG-PET} generation,'' {\em Biomedical Signal Processing and Control}, vol.~89, p.~105709, 3 2024.

\bibitem{Karimipourfard2023-gw}
M.~Karimipourfard, S.~Sina, F.~Khodadai~Shoshtari, and M.~Alavi, ``Synthesis of prospective multiple time points {F-18} {FDG} {PET} images from a single scan using a supervised generative adversarial network,'' {\em Nuklearmedizin}, vol.~62, pp.~61--72, Mar. 2023.

\bibitem{Takita2023}
H.~Takita, T.~Matsumoto, H.~Tatekawa, Y.~Katayama, K.~Nakajo, T.~Uda, Y.~Mitsuyama, S.~L. Walston, Y.~Miki, and D.~Ueda, ``{AI}-based virtual synthesis of methionine {PET} from contrast-enhanced {MRI}: Development and external validation study,'' {\em Radiology}, vol.~308, 8 2023.

\bibitem{Traverso2022}
A.~Traverso, C.~Rao, A.~Briassouli, A.~Dekker, D.~D. Ruysscher, and W.~van Elmpt, ``{PO-1609} generating synthetic hypoxia images from {FDG-PET} using generative adversarial networks ({GANs}),'' {\em Radiotherapy and Oncology}, vol.~170, pp.~S1396--S1397, 5 2022.

\bibitem{Nehmeh2021}
S.~A. Nehmeh, M.~B. Moussa, N.~Lee, P.~Zanzonico, M.~G{\"o}nen, J.~L. Humm, and H.~Sch{\"o}der, ``Comparison of {FDG and FMISO} uptakes and distributions in head and neck squamous cell cancer tumors,'' {\em EJNMMI Research}, vol.~11, p.~38, Apr 2021.

\bibitem{ACRIN}
P.~Kinahan, M.~Muzi, B.~Bialecki, and L.~Coombs, ``Data from {ACRIN-FMISO-Brain},'' 2018.

\bibitem{Rohlfing2010-zm}
T.~Rohlfing, N.~M. Zahr, E.~V. Sullivan, and A.~Pfefferbaum, ``The {SRI24} multichannel atlas of normal adult human brain structure,'' {\em Hum Brain Mapp}, vol.~31, pp.~798--819, May 2010.

\bibitem{Sun2015}
X.~Sun, L.~Shi, Y.~Luo, W.~Yang, H.~Li, P.~Liang, K.~Li, V.~C.~T. Mok, W.~C.~W. Chu, and D.~Wang, ``Histogram-based normalization technique on human brain magnetic resonance images from different acquisitions,'' {\em BioMedical Engineering OnLine}, vol.~14, July 2015.

\bibitem{K2013}
K.~K., {\em Basic PET Data Analysis Techniques}.
\newblock InTech, Dec. 2013.

\bibitem{Friston2011-jz}
K.~Friston, {\em Statistical parametric mapping}.
\newblock Academic Press, Apr. 2011.

\bibitem{gorgolewski_2016_50186}
K.~J. Gorgolewski~et al., ``{Nipype: a flexible, lightweight and extensible neuroimaging data processing framework in Python. 0.12.0-rc1},'' Apr. 2016.

\bibitem{Menze2015-fb}
B.~H. Menze~et al., ``The multimodal brain tumor image segmentation benchmark ({BRATS}),'' {\em IEEE Trans. Med. Imaging}, vol.~34, pp.~1993--2024, Oct. 2015.

\bibitem{Fedorov2012-qe}
A.~Fedorov, R.~Beichel, J.~Kalpathy-Cramer, J.~Finet, J.-C. Fillion-Robin, S.~Pujol, C.~Bauer, D.~Jennings, F.~Fennessy, M.~Sonka, J.~Buatti, S.~Aylward, J.~V. Miller, S.~Pieper, and R.~Kikinis, ``{3D} slicer as an image computing platform for the quantitative imaging network,'' {\em Magn. Reson. Imaging}, vol.~30, pp.~1323--1341, Nov. 2012.

\bibitem{Kamnitsas2017-rm}
K.~Kamnitsas, C.~Ledig, V.~F.~J. Newcombe, J.~P. Simpson, A.~D. Kane, D.~K. Menon, D.~Rueckert, and B.~Glocker, ``Efficient multi-scale {3D} {CNN} with fully connected {CRF} for accurate brain lesion segmentation,'' {\em Med. Image Anal.}, vol.~36, pp.~61--78, Feb. 2017.

\end{thebibliography}

\end{document}